\documentclass[12pt,a4paper]{article}
\usepackage{amsmath}
\usepackage{amssymb}
\usepackage{slashed}
\usepackage{color}
\usepackage[normalem]{ulem} 
\usepackage[normalem]{ulem}
\usepackage[pdftex]{graphicx}
\usepackage[colorlinks,citecolor=blue,urlcolor=blue,linkcolor=blue]{hyperref}
\usepackage[backend=bibtex,sorting=none,style=numeric-comp]{biblatex}

\newcommand{\GeV}{\mbox{$~{\rm GeV}$}}

\title{\bf \mbox{Direct detection of fermion dark matter}
\\ in the radiative seesaw model}
\author{Alejandro Ibarra\footnote{alejandro.ibarra@ph.tum.de} \\
\it \small Physik-Department T30d, Technische Universit\"at M\"unchen,\\ 
\it \small James-Franck-Stra{\ss}e, 85748 Garching, Germany,\\[4mm]
Carlos E. Yaguna\footnote{carlos.yaguna@mpi-hd.mpg.de} \\ 
\it\small Max-Planck-Institute f\"ur Kernphysik,\\
\it \small Saupfercheckweg 1, 69117 Heidelberg, Germany\\
\it \small and Institut f\"ur Theoretische Physik, Universit\"at M\"unster,\\
\it \small Wilhelm-Klemm-Stra\ss e 9, D-48149 M\"unster, Germany,\\[4mm]
Oscar Zapata\footnote{oalberto.zapata@udea.edu.co}\\
\it \small  Instituto de F\'{i}sica, Universidad de Antioquia,\\
\it \small  Calle 70 No. 52-21, Medell\'{i}n, Colombia}
\date{}

\begin{document}

\maketitle
\vspace*{-13cm}
\begin{flushright}
\texttt{\footnotesize TUM-HEP 967/14}\\
\texttt{\footnotesize FLAVOUR(267104)-ERC-87}\\
\texttt{\footnotesize MS-TP-14-36}
\end{flushright}
\vspace*{10cm}
\begin{abstract}
We consider the scenario in the radiative seesaw model where the dark matter particle is the lightest $Z_2$-odd fermion. We identify the regions of the parameter space of the model compatible with neutrino oscillation data, with the upper limits from rare charged lepton decays and with the observed dark matter abundance via thermal freeze-out, and we compute the dark matter scattering cross section with nuclei via the one-loop exchange of a photon, a $Z^0$-boson or a Higgs boson. We find that the predicted spin-independent cross section lies below the current LUX limit, although, for some choices of parameters, above the expected sensitivity of XENON1T or LZ. 
\end{abstract}

\section{Introduction}
The radiative seesaw model \cite{Ma:2006km}, or scotogenic model as is also known, is a simple extension of the Standard Model (SM) that can simultaneously account for neutrino masses and dark matter. It contains an extra scalar doublet, $H_2$, and at least two additional singlet fermions, $N_i$, $i=1,2...$,  all assumed to be odd under a $Z_2$ symmetry. In this model neutrino masses vanish at tree-level but are generated via quantum effects induced by the new fields. Furthermore, the lightest particle of the $Z_2$ odd sector, either the scalar doublet or the lightest fermion singlet, constitutes a dark matter candidate. 

The phenomenology of the scenario where the dark matter is the lightest fermion singlet has been extensively studied in recent years (see, {\it e.g.} \cite{Kubo:2006yx,Sierra:2008wj,Suematsu:2009ww,Gelmini:2009xd,Hambye:2009pw,Aoki:2010tf,Schmidt:2012yg,Ma:2012if,Hu:2012az,Kashiwase:2012xd,Kashiwase:2013uy,Racker:2013lua,Toma:2013zsa,Molinaro:2014lfa,Vicente:2014wga,Merle:2015gea,Merle:2015ica}), however the potential signals in direct detection experiments have received much less attention. In \cite{Schmidt:2012yg} it was assumed  that the two singlet fermions $N_1$ and $N_2$ are very degenerate, $\Delta M_N\sim \mathrm{keV}$, such that inelastic scattering on nuclei can take place. In that case, the dominant contribution to dark matter direct detection comes from a one-loop induced magnetic dipole operator, being the predicted rates sizable in some instances.  
This conclusion holds, however, only in this restricted framework. In general, $N_1$ and $N_2$ 
have a much larger mass splitting and therefore  the inelastic  scattering on nuclei is kinematically forbidden. 

The elastic scattering with nuclei is nonetheless possible, via the one-loop exchange of a photon,  a $Z$ boson and the Higgs boson.
While the detection of the radiatively-induced dark matter-nucleon interactions is challenging, the impressive current sensitivity of direct detection experiments, as well as their steady increase in reach, might allow to set significant limits on the parameters of the model, or optimistically, allow the observation of signals in the future. More specifically,  the current bound on the dark matter spin-independent interaction was set by the LUX experiment at the end of 2015, reaching a maximum value below $10^{-9}$ pb \cite{Akerib:2015rjg}. Future experiments are likely to improve the reach by almost three orders of magnitude \cite{Cushman:2013zza}. In fact, the XENON1T experiment \cite{Aprile:2012zx} recently started operation and it is expected to reach a  sensitivity of order $10^{-11}$ pb. In addition, the LZ experiment \cite{Malling:2011va,Kudryavtsev:2015vja}, the LUX successor, will start operating in 2018 and its expected sensitivity is of order $10^{-12}$ pb. 

When $N_1$ is the lightest $Z_2$ odd particle, the observed dark matter density can be generated through the freeze-out, superWIMP or freeze-in mechanisms, with either warm or cold dark matter \cite{Kubo:2006yx,Sierra:2008wj,Suematsu:2009ww,Gelmini:2009xd,Schmidt:2012yg,Ma:2012if,Hu:2012az,Molinaro:2014lfa}. In the present work we will focus on the freeze-out mechanism, which requires sizable Yukawa couplings between the dark matter and the Standard Model particles. As is well known, the same Yukawa couplings that keep the dark matter particles in thermal equilibrium at high temperatures, and that eventually allow their freeze-out, also induce neutrino masses and rare charged lepton decays. The flavor mixing observed in neutrino oscillation experiments requires the dark matter particle to couple to more then one leptonic mass eigenstate. As a result, the model predicts sizable rates for the leptonic rare decays, which are generically in tension with the stringent experimental upper limits. Some works have been devoted to find solutions to this tension, by imposing concrete flavor structures in the leptonic sector \cite{Kubo:2006yx,Suematsu:2009ww,Hu:2012az,Ho:2013hia,Ho:2013spa}.

In this paper, we perform a phenomenological analysis of the Yukawa sector of the model and we identify the regions of the parameter space compatible with the observed neutrino parameters, with the upper limits on rare leptonic decays, and with the generation of the observed dark matter abundance via thermal freeze-out. After identifying the viable parameter space of the model, we compute analytically the one-loop spin-dependent and spin-independent dark matter cross sections, as well as the anapole moment, and we confront the predictions of the model to the current experimental limits and to the expected sensitivity of future experiments. Notably, we not only find regions of the parameter space fulfilling all the experimental requirements without fine-tunings, but that for some choices of parameters the spin-independent cross section may indeed be large enough to produce signals at XENON1T or LZ.

The rest of the paper is organized as follows. In the next section, we review the radiative seesaw model and introduce our notation. In Section \ref{sec:meg} we  analyze the viable parameter space of the model, in particular we identify those scenarios leading to a suppressed rate for $\mu\rightarrow e\,\gamma$.  In section \ref{sec:dd} we calculate the interaction terms of the dark matter with a nucleus via the exchange of a photon, a $Z$-boson or the Higgs boson for parameters leading to the observed dark matter abundance via thermal freeze-out, and we confront the predictions of the model to the upper limits on the scattering rate from direct detection experiments. Finally, in section \ref{sec:con} we present our conclusions.

\section{The model}
\label{sec:mod}
The radiative seesaw model, also known as the scotogenic model \cite{Ma:2006km}, is a simple  extension of the Standard Model (SM) with a very rich phenomenology. The model contains one additional scalar doublet $H_2=(H^+,H_2^0)^T$ and at least two Majorana singlet fermions $N_i$ ($i=1,2,\ldots$), all assumed to have masses in the range between a few GeV and a few TeV. Furthermore, the model postulates that the vacuum displays an exact $Z_2$ symmetry under which the new fields are odd while the SM fields are even. This symmetry prevents tree-level charged lepton flavor violation and renders stable the lightest odd particle in the spectrum, which  becomes a dark matter candidate. In this model, the role of dark matter can be played by the neutral scalar or pseudo-scalar or by the lightest singlet fermion  \cite{Kubo:2006yx,Sierra:2008wj,Suematsu:2009ww,Gelmini:2009xd,Schmidt:2012yg,Kashiwase:2012xd,Kashiwase:2013uy,Klasen:2013jpa,Chakrabarty:2015yia}. In addition, this model can also account for neutrino masses \cite{Ma:2006km,Ma:2012if}, explain the baryon asymmetry of the Universe \cite{Hambye:2009pw,Racker:2013lua}, generate new signals at colliders \cite{Sierra:2008wj,Aoki:2010tf}, and induce observable rates for lepton flavor violating processes \cite{Toma:2013zsa,Vicente:2014wga}.

The Lagrangian of the model is:
\begin{align}
{\cal L}={\cal L}_\text{SM}+{\cal L}_{N_i}+{\cal L}_{H_2}+{\cal L}_\text{int},
\end{align}
where ${\cal L}_\text{SM}$ is the Standard Model Lagrangian, which includes the Higgs potential $V(H_1)=-\mu_{1}^{2}H_{1}^{\dagger}H_{1}+\lambda_{1}(H_{1}^{\dagger}H_{1})^{2}$. Besides, ${\cal L}_{N_i}$ and ${\cal L}_{H_2}$ contain, respectively, the terms involving only the $Z_2$-odd fermionic singlets $N_i$ and the scalar doublet $H_2$:
\begin{align}
{\cal L}_{N_i}&=\bar{N}_ii\slashed\partial P_R N_i- \frac{1}{2}M_i\left(\bar N_i^cP_RN_i+\mbox{h.c.}\right),\\
{\cal L}_{H_2}&=\left(D_\mu H_2\right)^\dagger\left(D^\mu H_2\right)-\mu_{2}^{2}H_{2}^{\dagger}H_{2}-\lambda_{2}\left(H_{2}^{\dagger}H_{2}\right)^{2}.
\end{align}
Here, and in the rest of the paper, we  choose to work in the basis where the $Z_2$-odd fermionic singlet mass matrix and the charged-lepton Yukawa matrix are diagonal. 
Finally, ${\cal L}_\text{int}$ contains the interaction terms between the $Z_2$-odd particles and the Standard Model particles:
\begin{align}
{\cal L}_\text{int}= -\left(Y_{\alpha i}\bar{L}_\alpha \tilde{H}_2 P_RN_i+\text{h.c.}\right)- V_\text{int}(H_1,H_2),
\end{align}
with an interaction potential identical to the one in the inert doublet model \cite{Deshpande:1977rw,Barbieri:2006dq,LopezHonorez:2006gr}
\begin{align}\label{eq:V}
V_\text{int}&=\lambda_{3}\left(H_{1}^{\dagger}H_{1}\right)\left(H_{2}^{\dagger}H_{2}\right)+\lambda_{4}\left(H_{1}^{\dagger}H_{2}\right)\left(H_{2}^{\dagger}H_{1}\right) +\frac{\lambda_{5}}{2}\left[\left(H_{1}^{\dagger}H_{2}\right)^{2}+{\rm h.c.}\right].
\end{align}
We demand that the mass terms in the scalar potential satisfy $\mu_{1}^2>0$, $2\lambda_1\mu_2^2> -\lambda_3\mu_1^2$, $2\lambda_1\mu_2^2> -(\lambda_3+\lambda_4\pm|\lambda_5|)\mu_1^2$ and that the quartic couplings fulfill  the vacuum stability conditions $\lambda_1,\,\lambda_2>0$ and $\lambda_3, \lambda_3+\lambda_4-|\lambda_5|>-2\sqrt{\lambda_1\lambda_2}$. We also require that only the $Z_2$-even scalar acquires a vacuum expectation value, $\langle H^0_1\rangle=v/\sqrt{2}$, with $v=246$ GeV, in order to render a vacuum also invariant under the $Z_2$ symmetry. Working in the unitary gauge, the scalar fields in the vacuum can be cast as:
\begin{equation} \label{eq:scalar-fields}
 H_1 =\frac{1}{\sqrt{2}}\begin{pmatrix} 0 \\  v+h\end{pmatrix} \;, \hspace{40pt} H_2= \begin{pmatrix} H^+ \\ \frac{1}{\sqrt2} \left( H^0 + i A^0 \right) \end{pmatrix}  \;,
\end{equation}
with $h$ the SM Higgs boson.  The masses of the new scalar particles are given by 
\begin{align}
m_{H^{0}}^{2}&=\mu_{2}^{2}+\frac{1}{2}\left(\lambda_{3}+\lambda_{4}+\lambda_{5}\right)v^2,\nonumber \\
m_{A^{0}}^{2}&=\mu_{2}^{2}+\frac{1}{2}\left(\lambda_{3}+\lambda_{4}-\lambda_{5}\right)v^2, \nonumber \\
m_{H^{\pm}}^{2}&=\mu_{2}^{2}+\frac{1}{2}\lambda_{3}v^2,	
\end{align}
which are constrained by current experiments to satisfy \cite{Pierce:2007ut,Lundstrom:2008ai}: $m_{A^0}+m_{H^0}>M_Z$, $m_{H^\pm}\gtrsim70$ GeV and max$[m_{A^0},m_{H^0}]\gtrsim100$ GeV. Besides, we will set the Standard Model Higgs mass to $126$ GeV \cite{Chatrchyan:2012ufa,Aad:2012tfa}.

The complete Lagrangian expressed in terms of the mass eigenstates $A^0,H^0, H^\pm, h$ is rather lengthy, therefore, here we will limit ourselves to write down only the interaction terms relevant for the dark matter phenomenology. These are the cubic interaction terms between two scalars and the Standard Model Higgs boson
\begin{align}
V\supset&\left[\frac{1}{2}\left(\lambda_3+\lambda_{4}+\lambda_{5}\right)(H^0)^2+\frac{1}{2}\left(\lambda_{3}+\lambda_{4}-\lambda_{5}\right)(A^0)^2+\lambda_{3}H^+H^-\right]\,v\,h,
\end{align}
the gauge interaction with the $Z$ boson
\begin{align}\nonumber
\mathcal{L}&\supset\frac{g}{2c_W}\left[i(1-2s^2_W)\left(H^+\partial^\mu H^--H^-\partial^\mu H^+\right)+\left(A^0\partial^\mu H^0-H^0\partial^\mu A^0\right)\right]Z^0_\mu,
\end{align}
with $c_W$ and $s_W$ the cosine and sine of the weak mixing angle, respectively,  and the Yukawa interaction of the scalars with the singlet fermions $N_i$
\begin{align}
{\cal L}\supset-Y_{\alpha i}\left(\frac{1}{\sqrt{2}}\bar{\nu}_\alpha(H^0-iA^0)-\bar{\ell}_\alpha H^-\right) P_RN_i+\text{h.c.},\label{eq:yuk}
\end{align}
where $\alpha=e,\mu,\tau$ denotes a leptonic flavor index and $i=1,2 ...$ runs over all fermionic singlets of the model.

One of the most notable features of the model is the generation at the one loop level of a neutrino mass term, induced  the Yukawa couplings $Y_{\alpha i}$. The neutrino mass matrix reads:
\begin{align}
  \left[\mathcal{M}_{\nu}\right]_{\alpha \beta}  &=\frac{\lambda_{5}v^{2}}{32\pi^{2}}\sum_{k}\frac{Y_{\alpha k}Y_{\beta k}}{M_k} \mathcal{I}\left(\frac{M_{k}^{2}}{m_{0}^{2}}\right),
\end{align}
with 
\begin{align}
\mathcal{I}(x)=\frac{x}{1-x}\left(1+\frac{x}{1-x}\log x\right).
\end{align}

We are interested in this paper in scenarios where the Yukawa couplings are sizable, typically $\gtrsim {\cal O}(0.1)$, and $m_0, M_k$ lie in the  range between a few GeV and a few TeV. Therefore, in order to generate neutrino masses smaller than $\sim 1$ eV, it is necessary a quartic coupling $|\lambda_5|\ll 1$. This requirement implies in particular $m_{H^0}\simeq m_{A^0}$, $m^2_{H^\pm}-m^2_{H^0}\simeq \lambda_4 v^2/2$. Furthermore, neutrino oscillation data restrict the flavor structure of the Yukawa couplings. To derive the most general form of the neutrino Yukawa matrix compatible with the neutrino oscillation data, we first cast the mass matrix as: 
\begin{align}
  \left[\mathcal{M}_{\nu}\right]_{\alpha \beta}  &=\sum_{k}Y_{\alpha k}Y_{\beta k}\tilde{M}_k=\left[Y\tilde{M}{Y}^T\right]_{\alpha\beta},
\label{eq:mass-matrix}
\end{align}
where $\tilde{M}=\text{diag}(\tilde{M}_1,\ldots,\tilde{M}_n)$ and
\begin{align} \label{eq:Mtilde}
\tilde{M}_k=\frac{\lambda_5v^2}{32\pi^2M_k}.
\end{align}
Finally,  following \cite{Casas:2001sr}, one obtains:
\begin{align}
Y=U^*\sqrt{m_\nu}R\sqrt{\tilde{M}^{-1}},
\label{eq:CI}
\end{align}
where  $U$ is the leptonic mixing matrix, $m_\nu=\text{diag}(m_{\nu_1},m_{\nu_ 2},m_{\nu_3})$, $m_{\nu_i}$ being the light neutrino masses, and $R$ is a complex orthogonal matrix, $R^TR=1$, of dimension $3\times n$, where $n$ is the number of singlet fermions.

In the freeze-out framework, the dark matter abundance is determined by the cross-sections of the various self-annihilation and coannihilation processes. Assuming that the dark matter and the scalar mediator masses are sufficiently non-degenerate coannihilations can be neglected, thus the relic abundance simply depends on the 
 total dark matter annihilation cross section; in this work we will take for concreteness $m_{H^+,H^0,A^0}\gtrsim 1.2 M_1$.
Expanding the cross section in powers of the relative dark matter velocity, $\sigma v =a +b v^2$, one finds
\begin{align} 
\Omega_{N_1} h^2 &\simeq \frac{1.07 \times 10^9\,{\rm GeV}^{-1}\, x_{\rm fo}}{g_*^{1/2}M_{\rm Pl}(a+3 b/x_{\rm fo})},
\end{align}
where $x_{fo}\equiv M_{1}/T_{fo}$, with $T_{fo}$ the temperature at which the freeze-out takes place, and  $g_*(x_{\rm fo})$ is the number of relativistic degrees of freedom in that epoch; for typical values of the parameters, $T_{\rm f.o.}\sim M_{1}/20$ and $g_*(x_{\rm f.o.})\sim 80$.

For the annihilation processes $N_1 N_1\rightarrow \ell_\alpha^+ \ell_\beta^-, \nu_\alpha \bar \nu_\beta$, and neglecting the lepton masses, one obtains \cite{Kubo:2006yx}
\begin{align}
a&=0, \nonumber \\
b&=\frac{M_1^2 y_1^4}{48\pi}\left[\frac{M_1^4+m_{H^+}^4}{(M^2_1+m^2_{H^+})^4}+\frac{M_1^4+m_{0}^4}{(M^2_1+m^2_{0})^4}\right],
\end{align}
where $y_1^2=\sum^3_{\alpha=1} |Y_{\alpha 1}|^2$ is the modulus squared of the column vector $Y_{\alpha 1}$. Therefore, the dark matter relic abundance can be estimated to be
\begin{align}\label{eq:omegaDM}
\Omega_{N_1} h^2&\simeq 0.12 \left(\frac{0.3}{y_1}\right)^4 \left(\frac{M_{1}}{100\GeV}\right)^2 \frac{(1+x)^4}{x^2(1+x^2)}\;.
\end{align}
Here, $x\equiv M_1^2/m_S^2$, where $m_S$ denotes a common mass for the scalar particles ($m_S=m_{H^+}=m_0$). 

In this model, the same Yukawa interactions that generate neutrino masses and the freeze-out of the dark matter particles also induce at the one-loop level lepton flavor violating processes, such as $\ell_\beta\to \ell_\alpha\,\gamma$ and $\mu-e$ conversion in nuclei. The decay branching ratio for the process $\ell_\beta\to\ell_\alpha\gamma$  reads \cite{Kubo:2006yx}
\begin{align}\label{eq:brmueg}
  \text{Br}(\ell_\beta\to \ell_\alpha\,\gamma) &=\frac{3\alpha_{\rm em}\text{Br}(\ell_\beta\to \ell_\alpha\nu_\beta\bar{\nu}_\alpha)}{64\pi G_F^2m_{H^\pm}^4} \left|\sum_k Y_{\alpha k}\, Y^{*}_{\beta k}\, F_{2}\left(\frac{M_{k}^{2}}{m_{H^{\pm}}^{2}}\right) \right|^{2},
\end{align}
where
\begin{align}
F_2(z)=\frac{1-6z+3z^2+2z^3-6z^2\ln z}{6(1-z)^4},
\end{align}
which approximately reads $F_2(z)\simeq 1/(3z)$ when $z\gg 1$ and $F_2(z)\simeq 1/6$ when $z \ll 1$. On the other hand, the $\mu-e$ conversion in nuclei reads~\cite{Toma:2013zsa,Vicente:2014wga}:
\begin{align}\label{eq:CR}
{\rm CR} (\mu- e, {\rm Nucleus}) &\approx\frac{2m_\mu^5 \alpha_{\mathrm{em}}^5Z_{\rm eff}^4F_p^2Z}{(4\pi)^4\Gamma_{\rm capt}m_{H^\pm}^4}|\sum_{k}Y_{ek}Y_{\mu k}^*H_2\left(\frac{M_{k}^{2}}{m_{H^{\pm}}^{2}}\right)|^2,
\end{align}
where $Z$ is the atomic number, $Z_{\rm eff}$ is the effective atomic charge,  $F_p$ is a nuclear matrix element and $\Gamma_{\rm capt}$ is the total muon capture rate; these are given, for different nuclei, in \cite{Kitano:2002mt,Vicente:2014wga}. Besides, $H_2(z)=\frac{1}{3}G_2(z)-F_2(z)$, with
\begin{eqnarray}
G_2(z) &=& \frac{2-9z+18z^2-11z^3+6z^3 \log z}{6(1-z)^4},
\end{eqnarray}
which has the asymptotic values $G_2(z)\simeq (-11/6+\log{z})/z$ for $z\gg 1$ and $G_2(z)\simeq 1/3$ for $z\ll 1$. 

The  current experimental limits Br$(\mu\to e\gamma)<5.7\times10^{-13}$ \cite{Adam:2013mnn}, Br$(\tau\to e\gamma)<3.3\times10^{-8}$ and Br$(\tau\to \mu\gamma)<4.4\times10^{-8}$ \cite{Aubert:2009ag}, $\text{CR}(\mu-e,\text{Ti})\leq 4.3\times10^{-12}$ \cite{Dohmen:1993mp}, $\text{CR}(\mu-e,\text{Au})\leq 7\times10^{-13}$ \cite{Bertl:2006up} set strong limits on the parameters of the model. In this paper we mainly focus on the rare decays $\ell_\beta\to \ell_\alpha\,\gamma$, which currently provide the most stringent limits on the model over most of the parameter space, however $\mu-e$ conversion in nuclei is likely to play also an important role in the future, due to the significant increase of sensitivity expected for this class of experiments \cite{Carey:2008zz,Glenzinski:2010zz,Abrams:2012er,Aoki:2010zz,Natori:2014yba,Cui:2009zz,Kuno:2013mha,Barlow:2011zza,PRIME}. Indeed, considering the generic case where $|Y_{e1}|\sim |Y_{\mu 1}|\sim |Y_{\tau 1}|\sim y_1/\sqrt{3}$, the expected branching ratio for $\mu\rightarrow e \gamma$ for the value of $y_1$ required by the freeze-out mechanism is approximately given by
\begin{align}
 \text{Br}(\mu \to e\,\gamma) \sim 4(2)\times 10^{-7} \left(\frac{M_{1}}{100\,{\rm GeV}}\right)^{-2}\;,
\end{align}
for $m_{H^\pm}/M_{1}=1.5 (10)$, which is orders of magnitude larger than the current upper limit, unless $M_{1}\gtrsim 100$ TeV, in tension with the requirement of partial wave unitarity for thermally produced dark matter~\cite{Griest:1989wd}. However, as it will be shown in the next section, there exist choices of parameters where the rate can be naturally suppressed, and hence lead to viable dark matter scenarios of thermal freeze-out.

\section{Scenarios with suppressed rates for $\mu\rightarrow e\gamma$}
\label{sec:meg}

We consider a scenario with the minimal particle content necessary to reproduce the two oscillation frequencies measured in neutrino experiments and incorporating a dark matter candidate, and which consists in adding to the Standard Model a $Z_2$-odd scalar doublet, $H_2$, and  two singlet $Z_2$-odd singlet fermions, $N_1$ and $N_2$.\footnote{The minimal extension is not unique, since the same features also arise by adding to the Standard Model particle content one $Z_2$-odd singlet fermion and two $Z_2$-odd scalar doublets~\cite{Hehn:2012kz}.} In this case, the orthogonal matrix $R$ defined in Eq.~(\ref{eq:CI}) reads \cite{Ibarra:2003up}:
\begin{align}\label{eq:nh}
R=\begin{pmatrix}
0 & 0\\
\cos\theta & -\sin\theta\\
\zeta \sin\theta & \zeta \cos\theta
\end{pmatrix},\hspace{1cm} \mbox{for normal hierarchy},
\end{align}
\begin{align}
R=\begin{pmatrix}
\cos\theta & -\sin\theta\\
\zeta\sin\theta & \zeta\cos\theta\\
0 & 0
\end{pmatrix},\hspace{1cm} \mbox{for inverted hierarchy}.
\end{align}
where $\theta$ is a complex parameter and $\zeta=\pm 1$ has been included to account for the possible reflections in the orthogonal matrix $R$.

We will focus in what follows in the case of normal hierarchy. Then, the Yukawa couplings explicitly read:
\begin{align}\label{eq:yuki1}
Y_{\alpha 1}&=\sqrt{\tilde{M}_1^{-1}}\left(\sqrt{m_{\nu_ 2}}\cos\theta U^{*}_{\alpha2} + \zeta\sqrt{m_{\nu_3}}\sin\theta U^{*}_{\alpha3}\right),\\\label{eq:yuki2}
Y_{\alpha2}&=\sqrt{\tilde{M}_2^{-1}}\left(-\sqrt{m_{\nu_ 2}}\sin\theta U^{*}_{\alpha2} + \zeta\sqrt{m_{\nu_3}}\cos\theta U^{*}_{\alpha3}\right)\;.
\end{align}
 These Yukawa couplings can be readily inserted in Eq.~(\ref{eq:brmueg}) to obtain the rates of the rare decays in terms of the low energy neutrino parameters, as well as the unknown parameters $\tilde M_k$, $m_{H^\pm}$ and the complex angle $\theta$. 

 To identify the scenarios with suppressed rate for $\mu\rightarrow e\gamma$ we first cast Eq. (\ref{eq:brmueg}) as:
\begin{align}\label{eq:brmueg_schem}
  \text{Br}(\mu \to e\,\gamma) \propto \left| Y_{e 1}\, Y^{*}_{\mu 1}\, F_{2}\left(\frac{M_{1}^{2}}{m_{H^{\pm}}^{2}}\right) +Y_{e 2}\, Y^{*}_{\mu 2}\, F_{2}\left(\frac{M_{2}^{2}}{m_{H^{\pm}}^{2}}\right) \right|^{2},
\end{align}
hence, this class of scenarios require, barring cancellations, especial choices of the neutrino parameters. More specifically, it is necessary that one of the terms in each addend is small, namely
\begin{center}
$Y_{\mu 1}\simeq 0 \quad {\rm or} \quad Y_{e 1}\simeq 0 $, \\ \vspace{0.1cm}
and \\ \vspace{0.1cm}
$\displaystyle{Y_{\mu 2}\simeq 0  \quad {\rm or} \quad Y_{e 2}\simeq 0 \quad {\rm or} \quad F_{2}\left(\frac{M_{2}^{2}}{m_{H^{\pm}}^{2}}\right)\simeq 0}$.
\end{center}
Notice that $F_2(M_1^2/m_{H^\pm}^2)>1/12$, since $M_1<m_{H^\pm}$, and therefore cannot be small, however, since the mass ordering between $M_2$ and $m_{H^\pm}$ is not fixed a priori, it is possible to find scenarios with $F_2(M_2^2/m_{H^\pm}^2)\simeq 0$ provided $M_2\gg m_{H^\pm}$. 

One can then identify scenarios with two texture zeros or one texture zero leading, without cancellations,  to a suppressed rate for $\mu\rightarrow e\,\gamma$  and which will lead to predictions for other observables. Moreover, there will be scenarios without texture zeros, also with suppressed $\text{Br}(\mu \to e\,\gamma)$, although with the price of cancellations among terms. Let us discuss separately these three possibilities.

\subsection{Two  texture zeros}

The simplest possibility to suppress the $\text{Br}(\mu\rightarrow e\gamma)$ consists on postulating two simultaneous zeros in the Yukawa matrices, namely: {\it i)} $Y_{e1}=0$, $Y_{e 2}=0$,  {\it ii)} $Y_{\mu 1}=0$, $Y_{\mu  2}=0$, {\it iii)} $Y_{e1}=0$, $Y_{\mu 2}=0$,  {\it iv)} $Y_{\mu 1}=0$, $Y_{e  2}=0$. Fixing two elements in the Yukawa matrix leads to a prediction on one of the low energy neutrino parameters~\cite{Frampton:2002qc,Ibarra:2003xp,Ibarra:2003up}. Concretely, the choice {\it i)} leads to $|U_{13}|\simeq 0.23$,  {\it ii)} to $|U_{13}|\simeq 0.25$,  and {\it iii)} and {\it iv)} to $|U_{13}|\simeq 0.08$, all of them outside the 3$\sigma$ range [0.126,0.178] obtained in ~\cite{Gonzalez-Garcia:2014bfa} from the global fit to the solar, atmospheric, reactor, and accelerator neutrino data. This scenario is therefore disfavored by present data.

\subsection{One texture zero}

A second possibility consists in postulating one texture zero, $Y_{\mu k}=0$ or $Y_{ek}=0$ for $k=1$ or $k=2$, and $F_2(M_l^2/m^2_{H^\pm})\simeq 0$ for $l\neq k$. Since we require $N_1$ to be the lightest $Z_2$-odd particle in the spectrum, this possibility can only be realized if the spectrum is of the form $M_1< m_{H^\pm}\ll M_2$ and either  $Y_{\mu 1}=0$ or $Y_{e 1}=0$. 

In the case $Y_{e1}=0$, the angle $\theta$ in the matrix $R$ reads:
\begin{align}
\tan\theta=-\sqrt{\frac{m_{\nu_ 2}}{m_{\nu_3}}}\frac{U^{*}_{e2}}{\zeta U^{*}_{e3}},
\end{align}
The flavor structure of the Yukawa couplings then becomes determined by the requirement of producing one texture zero and the low energy neutrino parameters $m_2^\nu$, $m_3^\nu$ and $U_{\alpha i}$. On the other hand, the overall size of the columns of the Yukawa matrices is determined by the free parameters $\tilde M_k$, which in turn depend on $\lambda_5$, the singlet masses and the heavy Higgs mass scale $m_0$ and thus have a rather obscure physical interpretation. Therefore, we prefer to cast the Yukawa couplings in terms of the modulus of the column vectors $y_1^2=\sum^3_{\alpha=1} |Y_{\alpha 1}|^2$, $y_2^2=\sum^3_{\alpha=1} |Y_{\alpha 2}|^2$. More specifically, we obtain 
\begin{align}
Y_{\alpha_1}&=\frac{y_1 e^{-i\omega/2}}{\sqrt{|U_{12}|^2+|U_{13}|^2}}
(U^*_{13}U^*_{\alpha 2}-U^*_{12}U^*_{\alpha 3}), \\
Y_{\alpha_2}&=\frac{y_2 e^{-i\omega/2}}{\sqrt{m^2_{\nu_2}|U_{12}|^2+m^2_{\nu_3} |U_{13}|^2}}(m_{\nu_2} U^*_{12}U^*_{\alpha 2}+m_{\nu_3} U^*_{13}U^*_{\alpha 3}),
\end{align}
with $\omega={\rm arg}(m_{\nu_ 2} U^{*2}_{12}+m_{\nu_3} U^{*2}_{13})$.

These expressions in turn allow to express the branching ratios of the rare leptonic decays only in terms of low energy neutrino parameter, the masses of the particles of the $Z_2$-odd sector and the overall size of the columns of the Yukawa coupling $y_1$, $y_2$. For the case $Y_{e1}=0$, the branching ratios for the electron-flavor  violating processes read:
\begin{align}
  \text{Br}(\mu \to e\,\gamma) &\simeq \frac{\alpha_{\rm em}y_2^4}{192\pi G_F^2 M_2^4}|f_{11}f_{12}|^2,\\
  \text{Br}(\tau \to e\,\gamma) &\simeq \frac{\alpha_{\rm em}y_2^4}{192\pi G_F^2 M_2^4}|f_{11}f_{13}|^2\text{Br}(\tau\to e \nu_\tau \bar{\nu}_e), \\
  \text{CR} (\mu- e, \text{Nucleus})&\simeq \frac{2m_\mu^5 \alpha_{\mathrm{em}}^5 y_2^4 Z_{\rm eff}^4F_p^2Z}{9(4\pi)^4\Gamma_{\rm capt}M_2^4}\left[\frac{17}{6}-\log\left(\frac{M_2^2}{m_{H^{\pm}}^{2}}\right)\right]^2 |f_{11}f_{12}|^2,
\end{align}
with
\begin{align} \label{eq:falphabeta}
f_{\alpha \beta}\equiv\frac{m_{\nu_2} U^*_{\alpha 2}U^*_{\beta 2}+m_{\nu_3} U^*_{\alpha 3}U^*_{\beta 3}}{\sqrt{m^2_{\nu_2}|U_{12}|^2+m^2_{\nu_3} |U_{13}|^2}},
\end{align}
while the branching ratio for the electron-flavor conserving decay reads:
\begin{align}\label{eq:efc}
  \text{Br}(\tau \to \mu\,\gamma) &\simeq\frac{\alpha_{\rm em}y_1^4}{768\pi G_F^2m_{H^\pm}^4}\frac{|U_{21}|^2 |U_{31}|^2}{(|U_{12}|^2+|U_{13}|^2)^2}\text{Br}(\tau\to \mu \nu_\tau \bar{\nu}_\mu).
\end{align}
In particular, in this scenario there is a correlation between the rates of $\tau\rightarrow e\,\gamma$ and $\mu\rightarrow e\,\gamma$ given just in terms of low energy neutrino parameters:
\begin{align}
\frac{\text{Br}(\tau \to e \,\gamma)}{\text{Br}(\mu \to e\,\gamma)}\simeq
\left|\frac{m_{\nu_2} U^*_{1 2}U^*_{3 2}+m_{\nu_3} U^*_{1 3}U^*_{3 3}}{m_{\nu_2} U^*_{1 2}U^*_{2 2}+m_{\nu_3} U^*_{1 3}U^*_{2 3}}\right|^2 \text{Br}(\tau\to e \nu_\tau \bar{\nu}_e)\sim 0.02-4.5,
\end{align}
the concrete value depending on the value of the phases $\delta$ and $\phi'$ in the leptonic mixing matrix. Moreover, lower limits on the masses of the next-to-lightest singlet and the charged Higgs states follow, respectively, from the present bound on $\mu\to e\,\gamma$ and $\tau\to \mu\,\gamma$:
\begin{align}
M_2&\gtrsim 2.4\, {\rm TeV}\left(\frac{y_2}{0.3}\right), \\
m_{H^\pm}&\gtrsim 120\, {\rm GeV}\left(\frac{y_1}{0.3}\right).
\end{align}
The best present limit on the model parameters from $\mu-e$ conversion stems from experiments with Ti  and Au, $\text{CR}(\mu-e,\text{Ti})\leq 4.3\times10^{-12}$, $\text{CR}(\mu-e,\text{Au})\leq 7\times10^{-13}$, and reads
\begin{align}
M_2&\gtrsim  0.6\, (1.0) \, {\rm TeV}\left(\frac{y_2}{0.3}\right)~~~\text{for Ti (Au)},
\end{align}
where we have taken for concreteness $M_2/m_{H^\pm}=10$. Given that the upper limit on $M_2$ scales as the fourth-power of the conversion rate, it would be necessary an increase in sensitivity by a factor $\sim 200$ in order to provide limits on the model competitive with those from $\mu\rightarrow e\gamma$; notably, this increase in sensitivity seems feasible in the future~\cite{PRIME}. It should be borne in mind, though, that the rate for $\mu-e$ conversion is suppressed when $17-6\log(M_2^2/m_{H^{\pm}}^{2})\approx 0$, thus the lower limit on $M_2$ from the non-detection of the process $\mu\rightarrow e \gamma$ is more robust, since cancellations among amplitudes never occur.

The discussion above only relies on the assumptions $Y_{e1}\simeq 0$ and $M_1< m_{H^\pm}\ll M_2$. Requiring that the dark matter is entirely produced via thermal freeze-out  leads to a lower limit on the dark matter mass. Indeed, taking the value of $y_1$ required by thermal production from Eq. (\ref{eq:omegaDM}), and substituting in  Eq. (\ref{eq:efc}), we obtain
\begin{align}
M_{1} &\gtrsim 140\,\text{GeV}\;.
\end{align}

An analogous calculation allows to obtain expressions for the Yukawa couplings in the case $Y_{\mu 1}=0$. The angle $\theta$ must take the value
\begin{align}
\tan\theta=-\sqrt{\frac{m_{\nu_ 2}}{m_{\nu_3}}}\frac{U^{*}_{\mu2}}{\zeta U^{*}_{\mu3}},
\end{align}
therefore,
\begin{align}
Y_{\alpha_1}&=\frac{y_1 e^{-i\omega//2}}{\sqrt{|U_{22}|^2+|U_{23}|^2}}
(U^*_{23}U^*_{\alpha 2}-U^*_{22}U^*_{\alpha 3}), \\
Y_{\alpha_2}&=\frac{y_2 e^{-i\omega/2}}{\sqrt{m^2_{\nu_2}|U_{22}|^2+m^2_{\nu_3} |U_{23}|^2}}(m_{\nu_2} U^*_{22}U^*_{\alpha 2}+m_{\nu_3} U^*_{23}U^*_{\alpha 3}),
\end{align}
with $\omega={\rm arg}(m_{\nu_ 2} U^{*2}_{22}+m_{\nu_3} U^{*2}_{23})$. The corresponding values for the branching ratios of the muon-flavor violating decays read:
\begin{align}
  \text{Br}(\mu \to e\,\gamma) &\simeq \frac{\alpha_{\rm em}y_2^4}{192\pi G_F^2 M_2^4}|g_{21}g_{22}|^2,\\
  \text{Br}(\tau \to \mu\,\gamma) &\simeq \frac{\alpha_{\rm em}y_2^4}{192\pi G_F^2 M_2^4}|g_{21}g_{23}|^2\text{Br}(\tau\to \mu \nu_\tau \bar{\nu}_e),\\
  \text{CR} (\mu- e, \text{Nucleus})&\simeq \frac{2m_\mu^5 \alpha_{\mathrm{em}}^5 y_2^4 Z_{\rm eff}^4F_p^2Z}{9(4\pi)^4\Gamma_{\rm capt}M_2^4}\left[\frac{17}{6}-\log\left(\frac{M_2^2}{m_{H^{\pm}}^{2}}\right)\right]^2 |g_{21}g_{22}|^2,
\end{align}
where $g_{\alpha\beta}$ is analogous to $f_{\alpha\beta}$, as defined in Eq.~(\ref{eq:falphabeta}), with the substitution $U_{1i} \rightarrow U_{2i}$, 
while the branching ratio for the muon flavor conserving decay is:
\begin{align}\label{eq:mufc}
  \text{Br}(\tau \to e\,\gamma) &\simeq\frac{\alpha_{\rm em}y_1^4}{768\pi G_F^2m_{H^\pm}^4}\frac{|U_{11}|^2 |U_{31}|^2}{(|U_{22}|^2+|U_{23}|^2)^2}\text{Br}(\tau\to e \nu_\tau \bar{\nu}_e).
\end{align}
This scenario leads to a correlation between the rates of $\tau \to \mu\,\gamma$ and $\mu \to e\,\gamma$:
\begin{align}
\frac{\text{Br}(\tau \to \mu \,\gamma)}{\text{Br}(\mu \to e\,\gamma)}\simeq
\left|\frac{m_{\nu_2} U^*_{2 2}U^*_{3 2}+m_{\nu_3} U^*_{2 3}U^*_{3 3}}{m_{\nu_2} U^*_{2 2}U^*_{1 2}+m_{\nu_3} U^*_{2 3}U^*_{1 3}}\right|^2 \text{Br}(\tau\to \mu \nu_\tau \bar{\nu}_\mu)\sim 1.4-28.8,
\end{align}
while the lower limits on the masses of the $Z_2$-odd sector from rare leptonic decays are:
\begin{align}
M_2&\gtrsim 2.1\, {\rm TeV}\left(\frac{y_2}{0.3}\right), \\
m_{H^\pm}&\gtrsim 120\, {\rm GeV}\left(\frac{y_1}{0.3}\right).
\end{align}
On the other hand, the most stringent current limit from $\mu-e$ conversion in nuclei gives
\begin{align}
M_2&\gtrsim 0.1\, (0.2) \, {\rm TeV}\left(\frac{y_2}{0.3}\right)~~~\text{for Ti (Au)},
\end{align}
again, for $M_2/m_{H^\pm}=10$.

As for the case $Y_{e1}\simeq 0$, also for this case one can set a lower limit on the dark matter mass from requiring thermal production. Using Eqs.~(\ref{eq:omegaDM}) and (\ref{eq:mufc}), we obtain
\begin{align}
M_{1} &\gtrsim 170\,\text{GeV}\,.
\end{align}

We conclude this subsection remarking that the scenarios with $M_1< m_{H^\pm}\ll M_2$ and $Y_{e1}=0$, or $Y_{\mu1}=0$, yield the smallest possible rates for $\mu\rightarrow e\gamma$ in the absence of fine-tunings. For any other choices of parameters, and barring cancellations,  the predicted rate for $\mu\rightarrow e\gamma$ will be larger and therefore the lower limits on the masses presented here can be regarded as conservative.

\subsection{No texture zero}

It is always possible to find a set of high energy parameters that leads to a suppressed rate for $\mu\to e\,\gamma$ by allowing cancellations among terms. This can be demonstrated noting first that $  \text{Br}(\ell_\beta\to \ell_\alpha\,\gamma) \propto |P_{\alpha\beta}|^2$, with
\begin{align}
P_{\alpha\beta }=\sum_k Y_{\alpha k} Y^*_{\beta k} F_{2}\left(x_k\right)
\label{eq:def-P}
\end{align}  
and $x_i=M_i^2/m_{H^\pm}^2$.
Defining
\begin{align}\label{eq:Ytilde}
Y'_{\alpha k}=Y_{\alpha k} \sqrt{F_{2}\left(x_k \right) }\;,
\end{align}
we have
\begin{align}\label{eq:P_tuning}
P_{\alpha\beta } =\sum_k Y'_{\alpha k} Y^{'*}_{\beta k} = (Y' Y'^\dagger)_{\alpha\beta}.
\end{align}
In terms of the rescaled Yukawa matrix $Y'_{\alpha k}$, the neutrino mass matrix Eq.~(\ref{eq:mass-matrix}) reads:
\begin{align}\label{eq:Mnu_tuning}
  \left[\mathcal{M}_{\nu}\right]_{\alpha \beta} =\sum_{k}Y'_{\alpha k}Y'_{\beta k}\tilde{M}'_k=\left[Y'\tilde{M'}{Y'}^T\right]_{\alpha\beta},
\end{align}
with
\begin{align}\label{eq:Mtilde2}
\tilde{M}'_k=\tilde{M}_k/F_{2}\left(x_k\right).  
\end{align}

Neutrino masses and the lepton flavor violating rates are then determined, respectively, by the symmetric matrix ${\cal M}_\nu= Y'\tilde{M}'{Y'}^T$ and by the Hermitian matrix $P=Y' Y'^\dagger$. As shown in \cite{Davidson:2001zk,Ibarra:2005qi}, for any $P$ and ${\cal M}_\nu$, it is possible to find matrices $Y'$ and $\tilde{M}'$ that solve Eqs.~(\ref{eq:P_tuning}) and ~(\ref{eq:Mnu_tuning}), and therefore, using Eqs.~(\ref{eq:Ytilde}) and (\ref{eq:Mtilde2}), matrices $Y$ and $\tilde{M}$.  In particular, there exists an infinite family of parameters leading to ${\rm Br}(\mu\rightarrow e\,\gamma)=0$, although most of these choices require a very large tuning among parameters.  Indeed, the present limits on rare decays require
\begin{align}
|P_{12}|&\lesssim 7\times 10^{-5}\left(\frac{m_{H^\pm}}{300\,\text{GeV}}\right)^{-2}, \nonumber \\
|P_{13}|&\lesssim 4\times 10^{-2}\left(\frac{m_{H^\pm}}{300\,\text{GeV}}\right)^{-2} , \nonumber \\
|P_{23}|&\lesssim 5\times 10^{-2}\left(\frac{m_{H^\pm}}{300\,\text{GeV}}\right)^{-2}.
\end{align}
Therefore, for Yukawa couplings of ${\cal O}(0.1)$, as required by thermal production, a fine cancellation among the different amplitudes contributing to the process $\mu\rightarrow e\gamma$ must take place in order to fulfill the stringent experimental upper limit in $|P_{12}|$, unless $m_H^\pm$ lies in the multi-TeV scale. 

In this scenario the  rate for $\mu\rightarrow e\gamma$ is suppressed by cancellations, however these cancellations do not necessarily occur simultaneously for other processes with $\mu-e$ flavor violation, potentially leading to signals in future experiments searching for $\mu-e$ conversion in nuclei or $\mu\rightarrow 3e$. More specifically, in the scenario where $\text{Br}(\mu\to e\,\gamma)$ vanishes due to cancellations,  the Yukawa couplings must be tuned to satisfy
\begin{align}
Y_{e2}Y_{\mu 2}^*=-Y_{e1}Y_{\mu 1}^*F_2\left(x_1\right)/F_2\left(x_2\right).
\end{align}
Substituting in Eq.~(\ref{eq:CR}) and taking for concreteness $Y_{\alpha 1}\sim y_1/\sqrt{3}$, the $\mu-e$ conversion rate takes the form 
\begin{align}
{\rm CR} (\mu- e, {\rm Nucleus}) &=\frac{m_\mu^5 \alpha_{\mathrm{em}}^5Z_{\rm eff}^4F_p^22Z}{\Gamma_{\rm capt}(4\pi)^4m_{H^\pm}^4}\frac{y_1^4}{9}\left|H_2(x_1)-H_2(x_2)F_2(x_1)/F_2(x_2)\right|^2.
\end{align}
For the mass spectrum $x_1\ll x_2\ll 1$ ($M_1^2<M_2^2\ll m_{H^\pm}^2$) we obtain
\begin{align}
{\rm CR} (\mu- e, {\rm Nucleus}) &\simeq\frac{m_\mu^5 \alpha_{\mathrm{em}}^5Z_{\rm eff}^4F_p^22Z}{\Gamma_{\rm capt}(4\pi)^4m_{H^\pm}^4}\frac{y_1^4}{9}\frac{x_2^2}{36}\\
&\approx1.9\,(1.6)\times10^{-17}\left(\frac{1\,\mbox{TeV}}{m_{H^\pm}}\right)^4\left(\frac{y_1}{0.3}\right)^4,~~~\text{for Ti (Au)}, 
\end{align}
where in the second line we have assumed $M_2/m_{H^\pm}=1/10$, for illustration. On the other hand, for the mass spectrum $x_1\ll1$ and $x_2\gg1$  ($M_1^2\ll m_{H^\pm}^2\ll M^2_2$)
\begin{align}
{\rm CR} (\mu- e, {\rm Nucleus}) &\simeq\frac{m_\mu^5 \alpha_{\mathrm{em}}^5Z_{\rm eff}^4F_p^22Z}{\Gamma_{\rm capt}(4\pi)^4m_{H^\pm}^4}\frac{y_1^4}{9}\frac{(\frac{5}{2}-\log x_2)^2}{36}\\
&\approx8.5\,(6.7)\times10^{-13}\left(\frac{1\,\mbox{TeV}}{m_{H^\pm}}\right)^4\left(\frac{y_1}{0.3}\right)^4,~~~\text{for Ti (Au)}, 
\end{align}
where we have taken here instead $M_2/m_{H^\pm}=10$. It is interesting that in the latter case the predicted rate for $\mu-e$ conversion is, for thermally produced dark matter, at the reach of projected experiments, provided $m_{H^\pm}\lesssim 1$ TeV; for the former case, however, the prospects for detection are poorer.
 
\section{Dark matter direct detection}
\label{sec:dd}
\begin{figure}[t!]
\begin{center}
\includegraphics[scale=0.7]{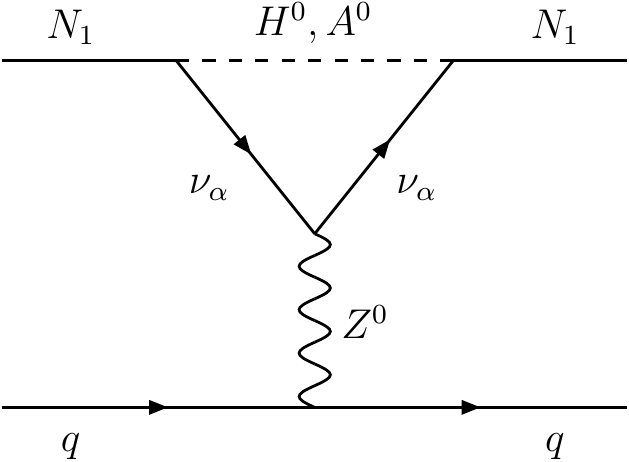} \hspace{1cm} 
\includegraphics[scale=0.7]{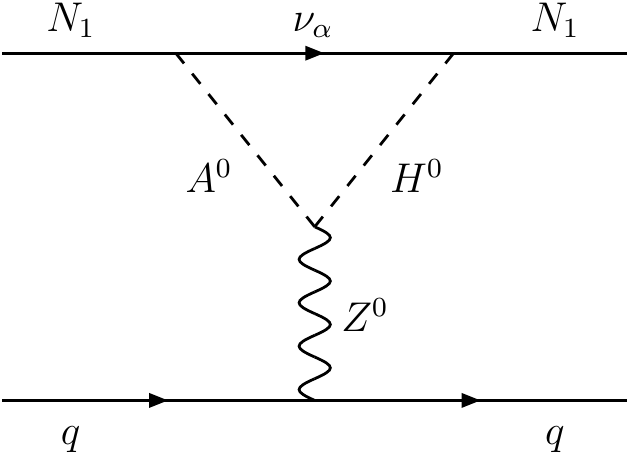} \\\vspace{0.5cm}
\includegraphics[scale=0.7]{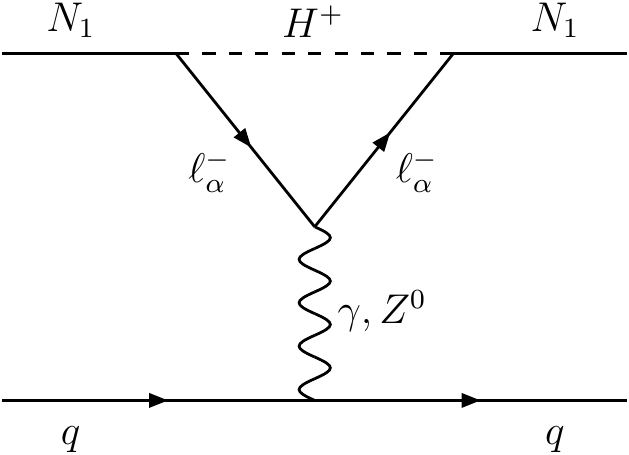} \hspace{1cm} 
\includegraphics[scale=0.7]{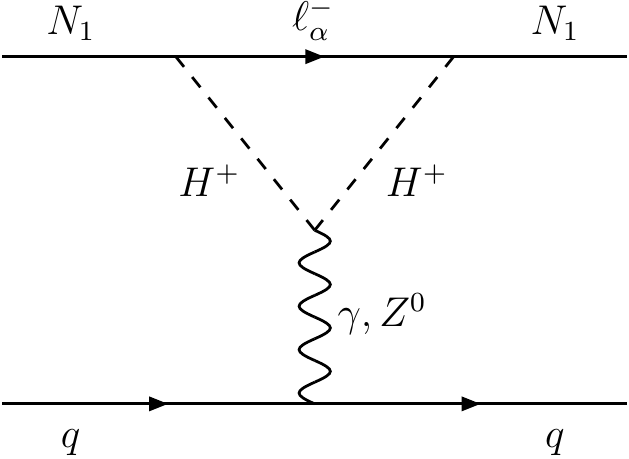} \\\vspace{0.5cm}
\includegraphics[scale=0.7]{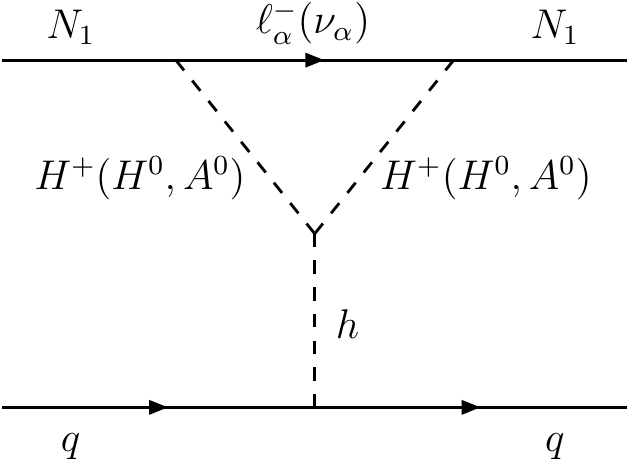}  \hspace{1cm}
\includegraphics[scale=0.7]{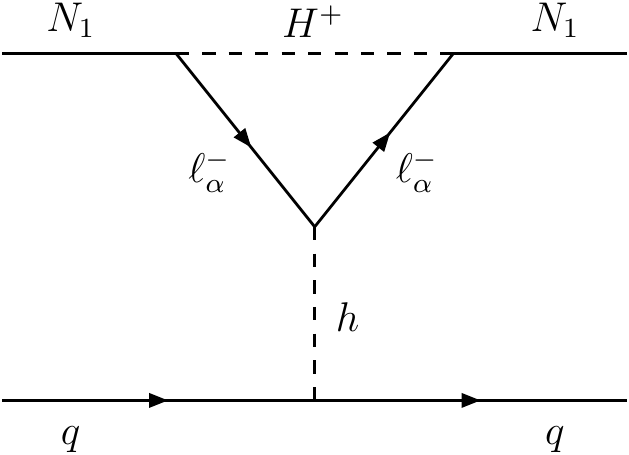}  
\caption{Diagrams contributing at one loop to the elastic scattering of dark matter particles off nuclei via the $Z$-boson exchange (first and second rows), photon exchange (second row) and  Higgs exchange (third row).}
\label{ddprocesses}
\end{center}
\end{figure}

Since the Yukawa term in Eq.~(\ref{eq:yuk}) is the only interaction between $N_1$ and the SM particles, quark-dark matter interactions do not occur at tree-level.  On the other hand, at the one-loop level, effective couplings do appear between $N_1$ and other SM particles, including the photon, the $Z$ boson and the Higgs boson (see Fig.~\ref{ddprocesses}). 

Let us start considering the scenario where the cubic interactions between the Higgs and the $Z_2$-odd scalars are very small, such that the dark matter coupling to the nucleon is dominated by the exchange of a gauge boson. The exchange of the $Z$-boson leads to an effective axial vector interaction term of the form $\xi_q\bar{N}_1\gamma^\mu\gamma^5N_1\bar{q}\gamma_\mu\gamma^5q$, where  $\xi_q$ reads:
\begin{align}
\label{eq:xiq}
\xi_q&=\frac{y_1^2a_q}{32\pi^2M_Z^2} \left[ (v_\ell+a_\ell)\mathcal{G}_2\left(\frac{M_1^2}{m_{H^\pm}^2}\right)+(v_\nu+a_\nu)\mathcal{G}_2\left(\frac{M_1^2}{m_{0}^2}\right)\right],
\end{align}
with $a_\ell=-\frac{g}{2c_W}\frac{1}{2},\,v_\ell=-\frac{g}{2c_W}(\frac{1}{2}-2s_W^2),\, v_\nu=a_\nu=\frac{g}{2c_W}\frac{1}{2}$ and $a_q=\frac{1}{2}(-\frac{1}{2})$ for  $q=u,c,t\, (d,s,b)$. The loop function $\mathcal{G}_2(x)$ is given by 
\begin{align}
\mathcal{G}_2(x)&=-1+\frac{2(x+(1-x)\ln(1-x))}{x^2},
\end{align}
and takes values between 0 and 1 for $0\leq x\leq 1$; for $x\lesssim0.1$ it can be well approximated by $x/3$. 
The resulting spin dependent cross section per nucleon $N$ is \cite{Jungman:1995df}
\begin{align}
\sigma_{N,SD}&=\frac{16}{\pi}\frac{M_1^2m_N^2}{(M_1+m_N)^2}J_N(J_N+1)\xi_N^2,
\end{align}
with  $\xi_N=\sum_{q=u,d,s}\Delta_q^N\xi_q$ and $\Delta_u^N=0.842,\,\Delta_d^N=-0.427,\Delta_s^N=-0.085$ \cite{Airapetian:2006vy}. The value of the spin-dependent cross section can then be estimated to be 
\begin{align}\nonumber
\label{eq:sd}
\sigma_{N,SD}^\mathrm{max}&\sim10^{-4}\mbox{pb}\left(\frac{y_1}{3.0}\right)^4
\mathcal{G}_2\left(\frac{M_1^2}{m_S^2}\right)^2\\
&\sim10^{-7}\mbox{pb}\left(\frac{M_1}{500\,\mbox{GeV}}\right)^2\mathcal{H}_2\left(\frac{M_1^2}{m_S^2}\right),
\end{align}
where in the second line we have substituted the value of $y_1$ necessary to generate the observed dark matter abundance via thermal freeze-out, Eq.~(\ref{eq:omegaDM}). Also, we have denoted $m_S$ as the common mass of all the extra scalar particles  and we have defined the function $\mathcal{H}_2(x)=(1+x)^4\mathcal{G}_2(x)^2/(x^2(1+x^2))$  which is monotonically increasing with $x$ and takes values between 1/9 and 8 for $0\leq x\leq 1$. 

\begin{figure}[t]
\begin{center} 
\includegraphics[scale=0.45]{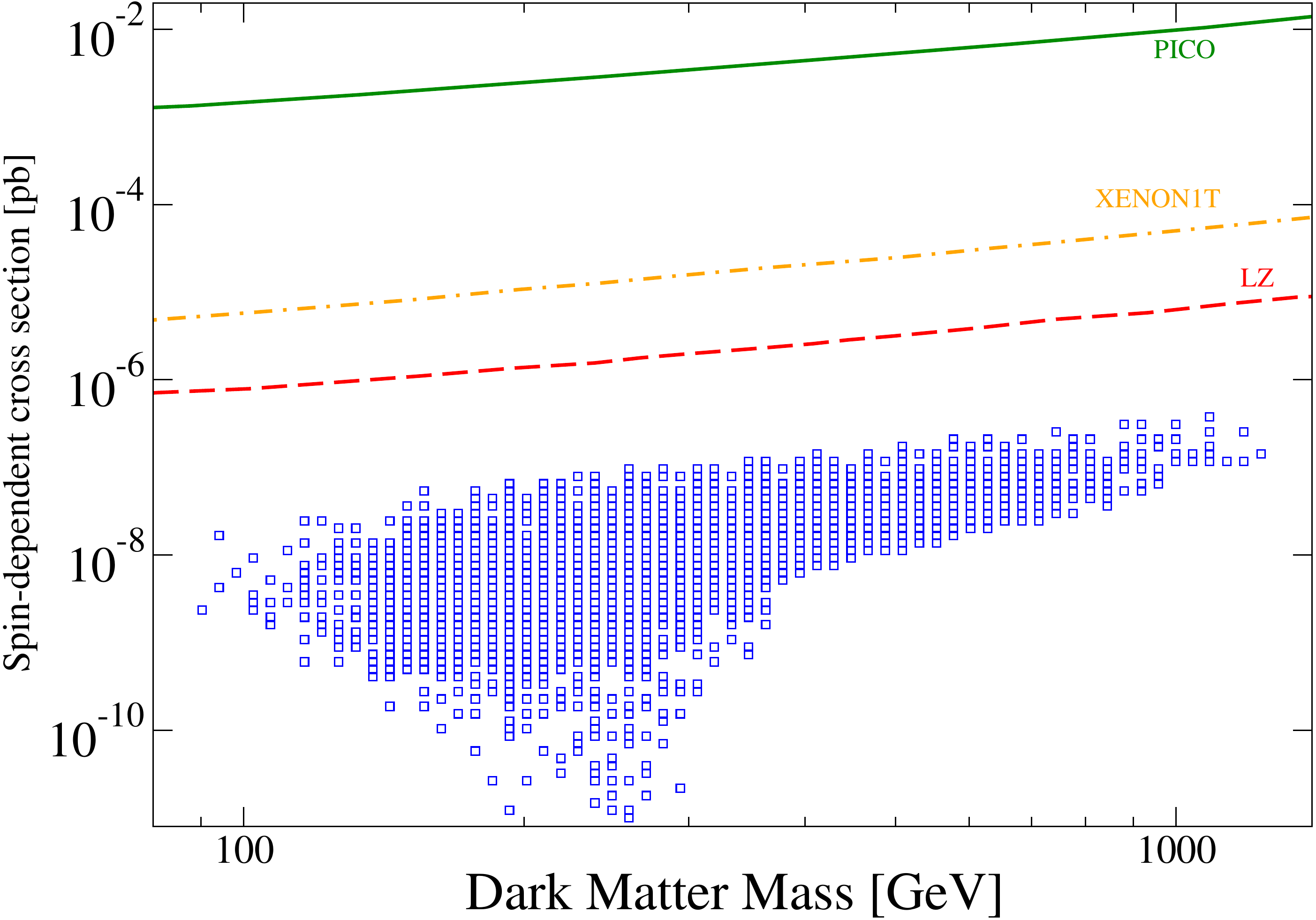}
\caption{\small Spin-dependent cross section as a function of the dark matter mass for a sample of viable points of the model, compared to the current upper limit from PICO (green solid  line) as well as to the projected sensitivities of XENON1T (yellow dot-dashed  line) and LZ (red dashed line). \label{fig:2gensd}}
\end{center}
\end{figure}

This estimate is confirmed by our numerical analysis, shown in Fig.~\ref{fig:2gensd}, where we calculate the spin-dependent cross section induced by the $Z$-boson exchange scanning over the relevant free parameters of the model as follows:
\begin{align}
1\,\text{GeV}&\leq M_1 \leq 3.5\,\text{TeV}, \nonumber  \\
M_1 &\leq M_2 \leq 10\,\text{TeV},  \nonumber   \\
1.2 M_1&\leq m_{H^0}, m_{H^\pm} \leq 5\,\text{TeV}.
\label{eq:ranges-scan}
\end{align}
All the points in the figure generate  a the dark matter density via thermal freeze-out in agreement with the value measured by Planck \cite{Ade:2013zuv}, which we calculate using micrOMEGAs~\cite{Belanger:2013oya}. For comparison, we also show in the figure the current bound on the spin-dependent cross section from PICO (green solid line) \cite{Amole:2015lsj} as well as the expected sensitivities of XENON1T (yellow dot-dashed line) and LZ (red dashed line) \cite{Cushman:2013zza}. As apparent from the plot, the predicted scattering rate induced by the exchange of the Z-boson lies well below the sensitivity of present and projected experiments. 

\begin{figure}[t]
\begin{center} 
\includegraphics[scale=0.45]{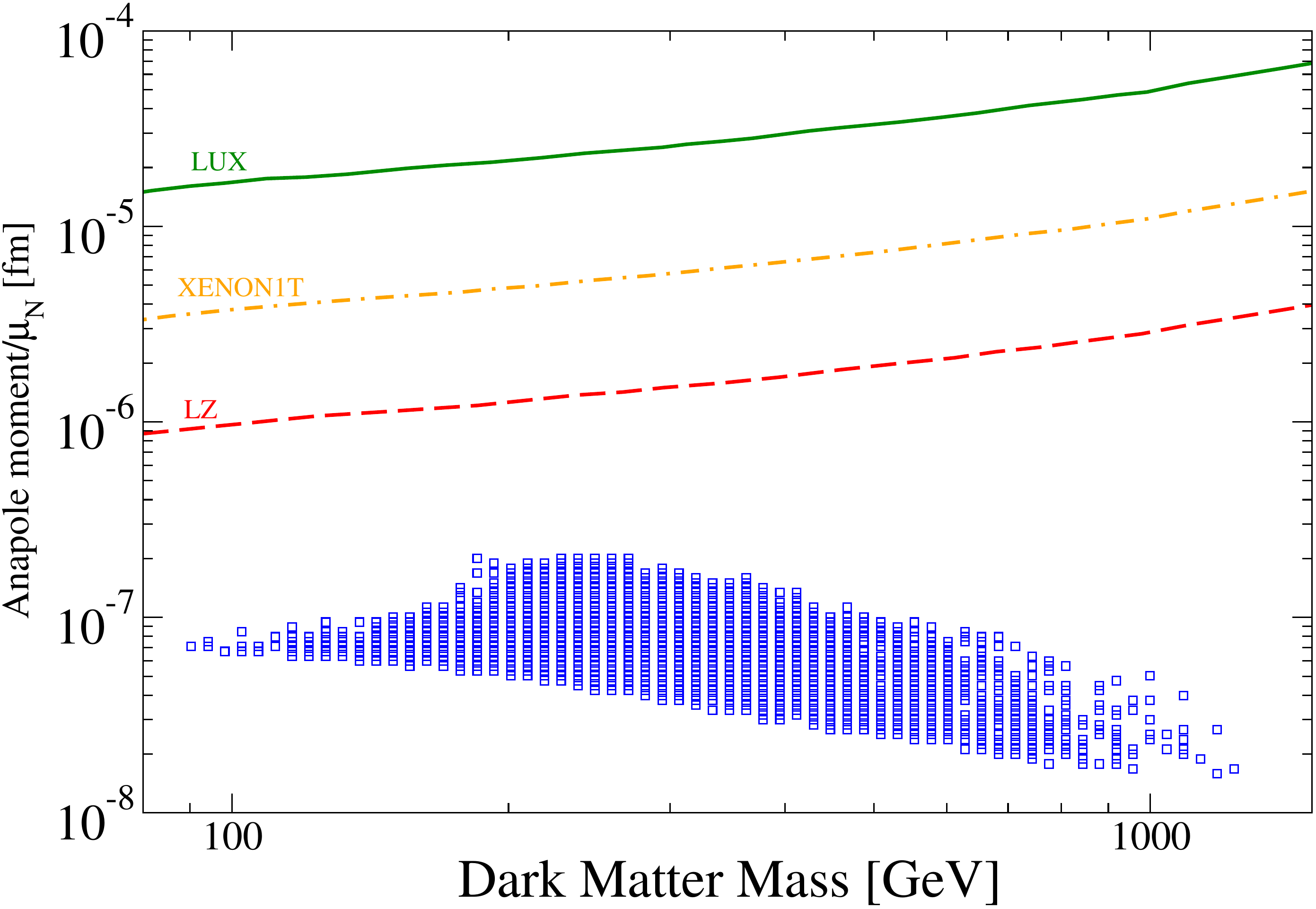}
\caption{\small Anapole moment as a function of the dark matter mass for a sample of viable points of the model, compared to the current upper limit from LUX (green solid  line) as well as to the projected sensitivities of XENON1T (yellow dot-dashed  line) and LZ (red dashed line). \label{fig:2genanapole}}
\end{center}
\end{figure}

The dark matter Yukawa interactions also give rise to an effective electromagnetic coupling at the quantum level.  Due to the invariance of the dark matter field under charge conjugation, the magnetic and electric dipole moment interactions identically vanish, and the leading contribution is the electromagnetic anapole moment~\cite{Ho:2012bg,Kopp:2014tsa,DelNobile:2014eta}:
\begin{align}
\mathcal{L}_{\mathcal{A}}=\mathcal{A}\bar{N}_1\gamma^\mu\gamma^5N_1\partial^\nu F_{\mu\nu}.
\end{align}
The anapole moment receives contributions from the exchange at one loop of each leptonic family, and reads \cite{Kopp:2014tsa}
\begin{align}
  {\cal A} &={\cal A}_e+{\cal A}_\mu+{\cal A}_\tau,  
\end{align}
where
\begin{align}
  {\cal A}_\ell &= -\frac{e|Y_{\ell1}|^2}{96\pi^2 M_1^2}\left[\frac{3}{2}\ln\left(\frac{\eta}{\epsilon}\right) - \frac{1+3\eta-3\epsilon}{\sqrt{(\eta-1-\epsilon)^2-4\epsilon}}\mbox{arctanh}\left(\frac{\sqrt{(\eta-1-\epsilon)^2-4\epsilon}}{\eta-1+\epsilon}\right) \right],
\end{align}
with $\eta=m^2_{H^\pm}/M_1^2$, $\epsilon=m^2_{\ell}/M_1^2$ and $\ell=\mu,\tau$. The electron contribution depends on the momentum transfer $q^2$ and is given by  
\begin{align}
  {\cal A}_e &= - \frac{e|Y_{e1}|^2}{32\pi^2 M_1^2}\left[\frac{-10+12\log\xi-(3+9\eta)\log(\eta-1)-(3-9\eta)\log\eta}{9(\eta-1)}\right],
\end{align}
with $\xi=|q^2|^{1/2}/M_1$ and $|q^2|\gg m_e^2$. 
We show in Fig.~\ref{fig:2genanapole} the predicted anapole moment for a sample of points of the parameter space of the model, confronted to the current limit by the LUX experiment (green solid  line) as well as to the projected sensitivities of XENON1T (yellow dot-dashed  line) and LZ (red dashed line) experiments; the predicted anapole moment lies well below the projected reach of projected experiments. We then conclude that the observation of signals in direct detection experiments will be challenging, if the dark matter interaction with the nucleon is dominated by the gauge interactions. 

\begin{figure}[t]
\begin{center} 
\includegraphics[scale=0.45]{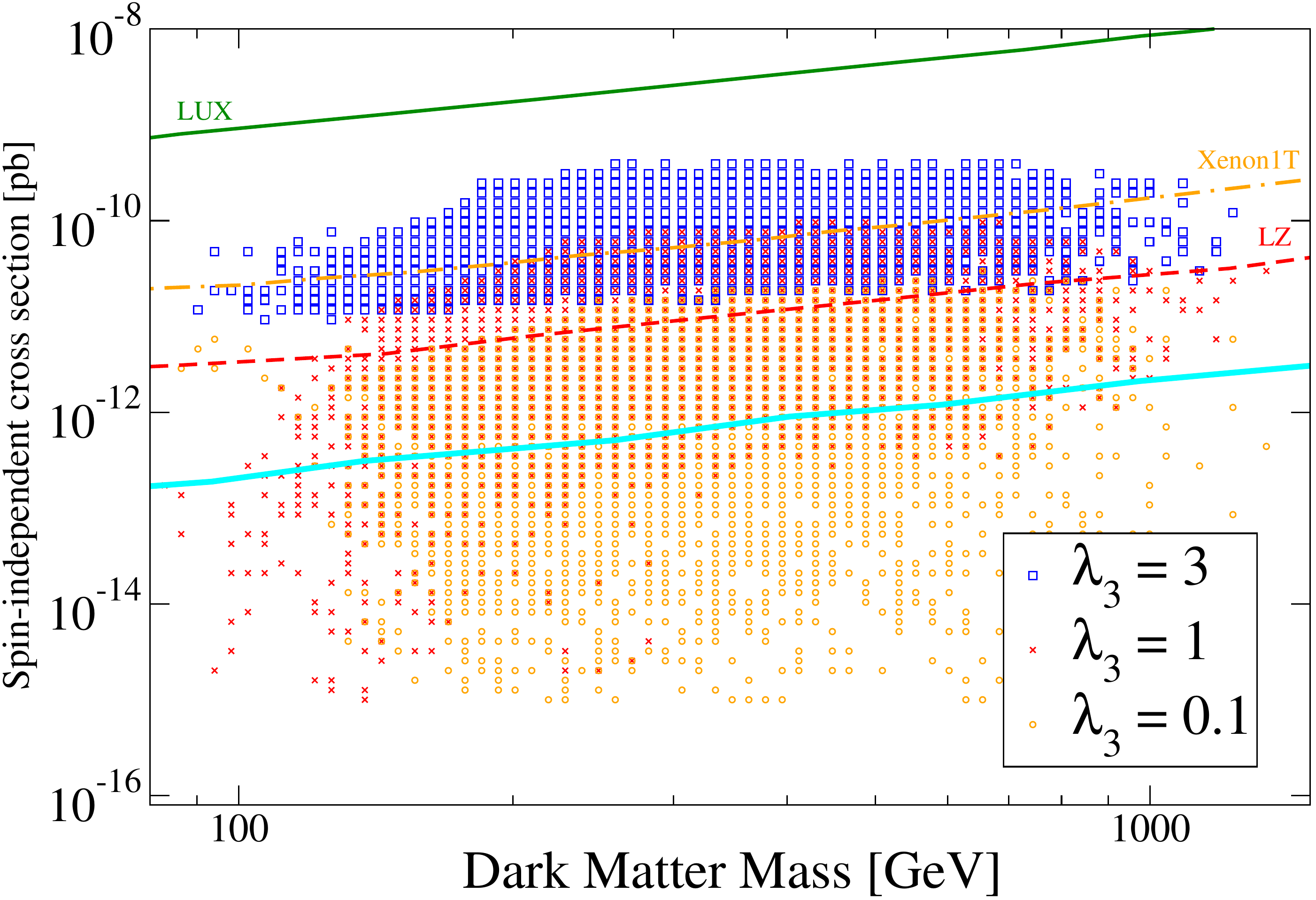}
\caption{\small  Spin-independent cross section as a function of the dark matter mass for a sample of viable points of the model, for different values of the coupling $\lambda_3$, compared to the current upper limit from LUX (green solid line) as well as to the projected sensitivities of XENON1T (yellow dot-dashed line) and LZ (red dashed line). The cyan line shows the neutrino coherent scattering limit.}
\label{fig:2gensip}
\end{center}
\end{figure}

The scattering rate of dark matter particles in a detector can be enhanced if the quartic couplings $\lambda_3$, $\lambda_4$ are sizable. If this is the case, the Higgs exchange diagrams shown in Fig.~\ref{ddprocesses} induce an effective scalar interaction term between $N_1$ and the quark $q$ of the form $\Lambda_{q}\bar{q}q\bar{N}_1N_1$, with  
\begin{align}
\label{eq:lambdaq}
\Lambda_q&=-\frac{y_1^2}{16\pi^2m_h^2M_1}\left[\lambda_3\mathcal{G}_1\left(\frac{M_1^2}{m_{H^\pm}^2}\right)+\frac{(\lambda_3+\lambda_4)}{2}\mathcal{G}_1\left(\frac{M_1^2}{m_{0}^2}\right)\right]m_q,
\end{align}
where $y_1=\left(\sum_\alpha |Y_{\alpha 1}|^2\right)^{1/2}$, and where we have neglected contributions proportional to the small charged lepton Yukawa couplings or to $\lambda_5$. 
The loop function $\mathcal{G}_1(x)$ is defined as
\begin{align}\label{eq:G1}
\mathcal{G}_1(x)&=\frac{x+(1-x)\ln(1-x)}{x},
\end{align}
 and takes values between 0 and 1 for $0\leq x\leq 1$; for $x\lesssim0.1$ it can be well approximated by $x/2$. This interaction leads to the spin-independent cross section $\sigma_{SI}$ of $N_1$ off a proton, which reads
\begin{align}
\sigma_{SI}&=\frac{4}{\pi}\frac{M_1^2m_p^2}{(M_1+m_p)^2}m_p^2\left(\frac{\Lambda_q}{m_q}\right)^2f_p^2,
\end{align}
where $m_p$ is the proton mass and $f_p\approx0.3$ is the scalar form factor. 
The value of $\sigma_{SI}$ is estimated to be
\begin{align}
\sigma_{SI}
&\sim5\times 10^{-7}\mbox{pb}\left(\frac{ \lambda_{3,4}}{3.0}\right)^2\left(\frac{y_1}{3.0}\right)^4\mathcal{G}_1\left(\frac{M_1^2}{m_{0}^2}\right)^2\left(\frac{100\,\mbox{GeV}}{M_1}\right)^2\\
&\sim 8\times 10^{-11}\mbox{pb}\left(\frac{\ \lambda_{3,4}}{3.0}\right)^2\mathcal{H}_1\left(\frac{M_1^2}{m_{0}^2}\right)
\end{align}
where, again,  $y_1$ was set by Eq. (\ref{eq:omegaDM}) and we have defined the function $\mathcal{H}_1(x)=(1+x)^4\mathcal{G}_1(x)^2/(x^2(1+x^2))$, which is monotonically increasing with $x$ and takes values between 1/4 and 8 for $0\leq x\leq 1$. Even for large quartic couplings, the predicted cross section lies about one order of magnitude below the current LUX limits, however future experiments might reach the necessary sensitivity to observe signals in this scenario. In Fig.~\ref{fig:2gensip} we show the expected spin-independent cross section  for $\lambda_3=0.1$ (yellow circles), $1$ (red crosses) and  $3$ (blue squares) from our scan of the parameter space, Eq.~(\ref{eq:ranges-scan}). For comparison, we also show in the figure the current bound from LUX \cite{Akerib:2015rjg} (black solid line), as well as the projected sensitivities of XENON1T (magenta dashed line) and LZ (green dashed line) \cite{Aprile:2012zx,Malling:2011va,Baer:2013ssa}. As apparent from the plot, future experiments will be able to probe the singlet fermion dark matter scenario of the radiative seesaw model, for sufficiently large values of the quartic coupling $\lambda_3$. On the other hand, points of the parameter space with small value of $\lambda_3$ lie below the neutrino coherent scattering limit, shown in the plot as a cyan line \cite{Billard:2013qya,Cushman:2013zza}.

\section{Conclusions}
\label{sec:con}

We have investigated prospects for direct dark matter detection in the radiative seesaw model in a scenario where the dark matter candidate is the lightest singlet fermion. Pursuing a phenomenological approach, we have identified the regions of the parameter space which lead to the observed neutrino oscillation parameters and to the measured dark matter abundance via thermal freeze-out, while satisfying the stringent experimental limits on  the rare charged lepton decays. Notably, we have found choices of parameters where the rate for the process $\mu\rightarrow e\gamma$ can be suppressed without invoking cancellations among the different amplitudes contributing to the process.

We have found that, for thermally produced dark matter particles, the direct detection rates mainly depend on the size of the quartic interaction terms between the $Z_2$-odd scalars and the Standard Model Higgs boson and on the masses of the new particles of the model, but are totally uncorrelated to flavor observables, such as oscillation parameter or the rates for the rare charged lepton decays. When the cubic interactions are very small, the scattering process is dominated by the one-loop exchange of a photon, through an anapole interaction, and of a $Z$-boson, leading to a spin-dependent scattering. Our analysis shows that the predicted anapole interaction and the spin-dependent scattering cross section unfortunately lie well below the sensitivity of present and foreseeable experiments. On the other hand, for sufficiently large quartic couplings, the scattering process is dominated by the exchange of the Higgs boson, leading to a spin-independent scattering. While present experiments are not yet probing this model, the upcoming XENON1T and LZ experiments will be able to probe points of the parameter space where the relevant quartic couplings are $\gtrsim {\cal O}(0.1)$.

\section*{Acknowledgments}
The work of A.I. has been partially supported by the DFG cluster of excellence ``Origin and Structure of the Universe'' and by the ERC Advanced Grant project ``FLAVOUR'' (267104), the work of C.Y. by the Max Planck Society in the project MANITOP, and the work of O.Z. by Sostenibilidad-UdeA, UdeA/CODI grant IN650CE and COLCIENCIAS through the grant number 111-556-934918.

\printbibliography

\end{document}